%% file: Template.tex
\documentclass{article}
\usepackage{spconf,amsmath,graphicx}
\PassOptionsToPackage{hyphens}{url}
\usepackage{hyperref}
\usepackage{arabtex}

\usepackage{multirow}
\title{End-to-end anti-spoofing with RawNet2}
\name{\begin{tabular}{c}Hemlata Tak$^1$, Jose Patino$^1$, Massimiliano Todisco$^1$, Andreas Nautsch$^1$ \\
Nicholas Evans$^1$ and {Anthony Larcher}$^2$\thanks{This work is partly supported by the ExTENSoR project funded by the
French Agence Nationale de la Recherche (ANR) and the VoicePersonae
project funded by ANR and the Japan Science and Technology Agency.}
\end{tabular}} 
\address{$^1$EURECOM, Sophia Antipolis, France\\
 $^2$LIUM - Université du Maine, Le Mans, France}
\begin{document}
%
\maketitle
\begin{abstract}
Spoofing countermeasures aim to protect automatic speaker verification systems from attempts to manipulate their reliability with the use of spoofed speech signals. While results from the most recent ASVspoof 2019 evaluation show great potential to detect most forms of attack, some continue to evade detection. This paper reports the first application of RawNet2 to anti-spoofing. RawNet2 ingests raw audio and has potential to learn cues that are not detectable using more traditional countermeasure solutions. We describe modifications made to the original RawNet2 architecture so that it can be applied to anti-spoofing. For A17 attacks, our RawNet2 systems results are the second-best reported, while the fusion of RawNet2 and baseline countermeasures gives the second-best results reported for the full ASVspoof 2019 logical access condition. Our results are reproducible with open source software. 
\end{abstract}


\section{Introduction}

\input introNew

\label{sec:intro}

\section{Previous work}

This section provides a brief review of past work that led to the development of RawNet2~\cite{jung2020improved}. The treatment is upon automatic speaker verification. 
As in many fields of speech processing, recent years have seen the adoption of end-to-end classifier architectures where every component between, and including any front-end processing and back-end classification, is automatically and jointly optimised. Many of these solutions operate directly upon the raw speech waveform and avoid limitations introduced from the use of knowledge-based, hand-crafted acoustic features. One advantage is the use of representations optimised for the application, rather than the use of generic, fixed-bandwidth decompositions such as those stemming from Fourier based analysis~\cite{muckenhirn2018towards}.

There is already a sizeable body of work in the automatic speaker verification literature which reports end-to-end solutions that operate on the raw speech waveform. Among other early solutions is RawNet~\cite{jung2019rawnet}, introduced in 2019. RawNet is a convolutional neural network architecture which outputs speaker embeddings. The first convolutional layer is applied directly to the raw speech waveform, with all filter parameters being learned automatically. Among the higher layers are residual blocks~\cite{he2016deep} which extract frame-level representations. Residual blocks use skip connections that enable the training of deeper classifiers to leverage more discriminative information. They use either long short-term memory (LSTM) as in~\cite{jung2018complete} or gated recurrent units (GRUs) as in~\cite{jung2019rawnet} to aggregate utterance-level representations and either a b-vector classifier~\cite{lee2014speaker} as in~\cite{jung2018complete} or a DNN backend with concatenation and multiplication (concat $\&$ mul) technique as in~\cite{jung2019rawnet} for verification. The use of a wholly unconstrained first layer whose parameters are learned automatically can result in slow learning. The first layer outputs also tend to be noisy, especially when training data is sparse.

One solution to these issues is SincNet~\cite{ravanelli2018speaker,ravanelli2018learning}. Whereas the higher layers of the SincNet architecture are relatively standard, the first convolutional layer operates upon the raw waveform and consists in a bank of band-pass filters parametrized in the form of sinc functions. The use of a constrained first layer, with fewer learnable parameters whereby only the cut-in and cut-off frequencies are learned with a fixed rectangular-shaped filter response, leads to the learning of a more meaningful filterbank structure and outputs. 

RawNet2~\cite{jung2020improved}, proposed in 2020, combines the merits of the original RawNet approach (RawNet1) with those of SincNet. The first layer of RawNet2 is essentially the same as that of SincNet, whereas the upper layers consist of the same residual blocks and GRU layer as RawNet1. New to RawNet2 is the application of filter-wise feature map scaling (FMS) using a sigmoid function applied to residual block outputs as in~\cite{woo2018cbam}. FMS acts as an attention mechanism and has the goal of deriving more discriminative representations. The embedding dimension for RawNet2 is also greatly increased, from 128 for RawNet1 to 1024 for RawNet2.
Last, whereas RawNet1 obtains better results with a DNN-based back-end classifier, RawNet2 gives better results with a cosine similarity score.

Experiments using the VoxCeleb1 database show that
RawNet1 gives an 11\% relative improvement in terms of an equal error rate (EER) metric compared to an i-vector based system~\cite{shon2018frame}, 
and is also shown to be competitive with those of an x-vector system~\cite{okabe2018attentive}.
When trained using the VoxCeleb2 database, RawNet2 gives a 20\% relative improvement compared to an x-vector based system~\cite{snyder2018x} whereas, using the expanded VoxCeleb1-E evaluation dataset protocol, it gives a 13\% relative improvement compared to a Thin-ResNet-34 based system~\cite{nagrani2020voxceleb}. There is hence plentiful evidence that end-to-end architectures which avoid the use of knowledge-based, hand-crafted features have potential to improve automatic speaker verification performance. The goal of the work reported in this paper is to determine whether the benefit translates also to anti-spoofing, especially in worst case scenarios.

\section{Application to anti-spoofing}
This section describes modifications made to the original RawNet2 architecture~\cite{jung2020improved} so that it can be applied successfully to anti-spoofing. Modifications to parameters or architecture components are highlighted in Table~\ref{Tab:RawNet details} in bold text.

The first modification concerns the first layer of the architecture which ingests raw speech. Since it leads to worse performance, we did not apply layer normalization~\cite{ba2016layer} to the input. On account of training data sparsity, or rather the limited number of different spoofing attacks (only 6 for the training \& development partitions of the ASVspoof 2019 LA database), we neither learn automatically the bandwidth nor spectral position of each sinc filter. Other experiments not reported here show that it leads to overfitting. 
While SincNet is initialised with a Mel-distributed filterbank, we experimented also with linearly-distributed and inverse-Mel scaled filterbank. These choices are motivated by the success of linear and inverse-Mel~\cite{sahidullah2015comparison} front-ends for anti-spoofing. We fixed the duration of the raw waveform input to {$\approx 4$} sec (64000 samples) by either cropping long utterances or concatenating short utterances such that all utterances have the same length.

In similar fashion to~\cite{jung2020improved}, we also optimised the filter length (impulse response duration). This is because the duration of cues used to detect spoofing are not necessarily the same as those for speaker recognition. Whereas the original work used filter lengths of 251 samples, we observed better results using only 129 samples.
Whereas the configuration of the first residual block is left unchanged, we use a larger number of kernel filters (512) in second residual block. As for RawNet2, we apply FMS independently to the output of each residual-block, thereby helping to emphasize the most informative filter outputs. We adopt the combined additive and multiplicative feature scaling approach in~\cite{jung2020improved}.

A GRU layer with 1024 hidden nodes is applied to aggregate frame-level representations into a single utterance-level representation. Rather than producing embeddings as in RawNet 2, the GRU output is followed by an additional fully connected layer which precedes the output layer. A softmax activation function is then applied in order to produce two-class predictions: bona-fide or spoof. The network was trained with ADAM optimisation using a learning rate of 0.0001, 100 epochs and a mini-batch size of 32. The full architecture is summarised in Table~\ref{Tab:RawNet details}. 

\label{sec:typestyle}

 \begin{table}[!t]
\small
	\centering

	\caption{The RawNet2 architecture used for anti-spoofing. Modifications made to the original architecture are highlighted in boldface. BN refers to batch normalisation.}
 \setlength\tabcolsep{3.5pt}
	\begin{tabular}{ *{3}{c}}
	\hline
	Layer & Input:{\bfseries 64000} samples & Output shape\\
	\hline
	& Conv(\textbf{129},1,128)& \\
	\textbf{Fixed} Sinc filters&Maxpooling(3)&({\bfseries 21290},128)\\
	& BN \& LeakyReLU&\\
	\hline
	
	Res block & 
	    
	{$ \left \{ 
	\begin{array}{c}
	\text{BN  \&  LeakyReLU} \\
	   \text{Conv(3,1,128)} \\
	  \text{BN  \& LeakyReLU} \\
	  \text{Conv(3,1,128)}\\
	   \text{Maxpooling(3)}\\ 
	   \text{FMS}
	\end{array} \right \} 
	  
	  \times 2$} & ({\bfseries 2365},128) \\
	\hline
  Res block & 
  {$ \left \{ 
	\begin{array}{c}
	\text{BN  \&  LeakyReLU} \\
	   \text{Conv(3,1, \textbf{512})} \\
	  \text{BN \& LeakyReLU} \\
	  \text{Conv(3,1, \textbf{512})}\\
	   \text{Maxpooling(3)}\\ 
	   \text{FMS}
	\end{array} 
	 \right \} \times 4$} &({\bfseries 29},512) \\
	 \hline
	 GRU & GRU(1024)&(1024)\\
	 \hline
	 FC&1024&(1024)\\
	 \hline
	 \textbf{Output}&\textbf{1024}&\textbf{2}\\
	 \hline
	\end{tabular}
	\label{Tab:RawNet details}
	\vspace{-0.15cm}
\end{table}

\section{Experimental work}
\label{sec:Experimental-set}
The following describes the ASVspoof 2019 LA database, evaluation metrics, baseline countermeasure and results. We report results for three different RawNet2 variants and results for fused systems.

\subsection{ASVspoof 2019 and t-DCF metric}
The ASVspoof 2019 logical access (LA) database consists of three independent partitions: train, development and evaluation. Spoofed attacks are generated using a set of 19 different speech synthesis, voice conversion and hybrid algorithms~\cite{wang2020asvspoof}. The training and development partitions contain bona fide and spoofed data generated with 6 different attack algorithms (A01-A06), whereas the evaluation partitions contains bona fide speech and spoofed data generated with 13 different algorithms (A07-A19). We re-partitioned the data by supplementing training data with 90 \% of development data. The remaining 10\% was used for validation. The primary metric is the minimum normalised tandem detection cost function (t-DCF) metric~\cite{Kinnunen2018-tDCF,kinnunen-tDCF-TASLP}. For consistency with results reported in~\cite{todisco2019asvspoof}, t-DCF results derived according to the original min t-DCF metric in~\cite{Kinnunen2018-tDCF,kinnunen-tDCF-TASLP}.
Results are also reported in terms of the pooled equal error rate (EER) computed using~\cite{brummer2013bosaris}.

\subsection{Baseline}

The baseline classifier for this work is the high-spectral resolution linear frequency cepstral coefficient (LFCC) countermeasure reported in~\cite{IS2020LFCC}. It uses $70$ linearly-spaced filters with conventional cepstral analysis and a GMM back-end classifier. Despite its simplicity, it outperforms all but three systems submitted to the ASVspoof 2019 challenge.

\subsection{RawNet2 results}
\label{ssec:subhead}

\begin{table*}[!ht]
	\small
	\centering
	
	\caption{Results in terms of min t-DCF for development (A01-A06) and evaluation (A07-A19) partitions and respective pooled min~t-DCF (P1) and pooled EER (P2). L: High-spectral-resolution LFCC. 
	S1: fixed Mel-scaled sinc filters.
	S2: fixed inverse Mel-scaled sinc filters.
	S3: fixed linear-scale sinc filters.}
	\setlength\tabcolsep{0.7pt}
	\begin{tabular}{c || *{6}{c} |cc|| *{13}{c}| cc} 
		\hline
		
		S &A01&A02&A03&A04&A05&A06&P1&P2&A07&A08&A09&A10&A11&A12&A13&A14&A15&A16&A17&A18 &A19& P1&P2 	\\ 
		\hline\hline
		L&.0000&.0000&.0000&.0005&.0000&.0000&.0000&.00& .0011&.0009&.0003&.1536&.0048&.1150&.0798&.0697&.0695&.0007&.3524&.0741&.0084 	&\textbf{.0904}&	\textbf{3.50}\\
	
		\hline
		 S1&.0443&.0125&.0274&.0435&.0598&.0653&.0460&1.36&.0277&.1984&.0118&.0373&.0195&.0379&.0433&.0192&.0314&.0457&.2620&.6145&.0526&.1301&5.64\\
		 \hline
		
		S2&.0269&.0025&.0045&.0291&.0297&.0347&.0285&0.86&.0131&.0551&.0131&.0300&.0100&.0360&.0076&.0046&.0344&.0187&.2626&.6237&.0409&.1175&5.13
		\\
		\hline

		 S3&.0230&.0315&.0117 &.0468 & .0354& .0690 &.0400 &1.25 &.0382& .1106& .0348&.0493 &.0505&.0445 &.0459 &.0380 &.0378 &.0460 &\textbf{.1810}&.5285 &.0679&.1294&4.66\\
		\hline
	
	\end{tabular}
	\label{Tab: results on dev and eval}
\end{table*}

Table~\ref{Tab: results on dev and eval} shows results in terms of min t-DCF for the baseline~(L) and three different RawNet configurations~(S1-3). Attacks A01-06 correspond to development data whereas A07-A19 correspond to evaluation data. P1 and P2 columns show pooled min t-DCF and EER results respectively for each partition.

Results for all three RawNet classifiers are inferior to those of the baseline. Focusing only upon the evaluation set, the pooled min t-DCF for the baseline is 0.09, whereas the best performing S2 RawNet2 system which uses fixed inverse Mel-scaled sinc filters gives a min t-DCF of 0.1175. Results for the infamous A17 attack, i.e.\ the worst case for the baseline and all of the top-performing ASVspoof 2019 LA submissions, are more favourable. Whereas the baseline classifier gives a min t-DCF of 0.3524 for A17, the S3 RawNet2 classifier that uses fixed, linearly-scaled sinc filters gives a min t-DCF of 0.181. To the best of our knowledge, this is among very best published results for A17.

\subsection{Fusion}

\begin{table}[!t]
	\centering
	\small

	\caption{Performance for the ASVspoof 2019 evaluation partition in terms of pooled min t-DCF and pooled EER for top-performing systems (T05, T45, T60 and T24), SVM-based fusions of high-spectral-resolution LFCC (L)~\cite{IS2020LFCC} and RawNet2 systems (boldface), and ASVspoof 2019 baseline systems (B1, B2). Individual min t-DCF results for A17 are illustrated in the right-most column.}
 \setlength\tabcolsep{2.5pt}
	\begin{tabular}{ *{4}{c}}
		\hline
		 System & min-tDCF& EER& min t-DCF (A17) 	\\ 
	\hline	
	   T05&0.0069&0.22&0.0040\\
	   \hline
	  
	   {\bfseries L+S1}&{\bfseries 0.0330}&{\bfseries 1.12}&{\bfseries 0.1161}\\
	   \hline
	   {\bfseries L+S1+S2+S3} & {\bfseries 0.0347}&{\bfseries 1.14}&{\bfseries 0.0808}\\
	   
	   \hline
	   {\bfseries L+S3}&{\bfseries 0.0370}&{\bfseries 1.14}&{\bfseries 0.0965}\\
	   \hline
	  
	   { \bfseries L+S2}&{\bfseries 0.0443}&{\bfseries 1.35}&{\bfseries 0.1339}\\
	   \hline

	   T45~\cite{lavrentyeva2019stc}&0.0510&1.86&0.2208\\

		 \hline
		 
		  T60~\cite{chettri2019ensemble}&0.0755&2.64&0.3254\\
		\hline

		 L~\cite{IS2020LFCC} &0.0904&3.50&0.3524\\
		 \hline
		 T24& 0.0953 &3.45&0.3547\\ 
		 \hline

		 LFCC:B2~\cite{todisco2019asvspoof}&0.2116&8.09&0.2042\\
		 \hline
	   CQCC:B1~\cite{todisco2019asvspoof}&0.2366&9.57&0.5859\\
		\hline
	
	\end{tabular}
	\label{Tab:comparsion results}
	\vspace{-0.15cm}
\end{table}

Since the LFCC-GMM baseline system gives the best over-all pooled result, whereas the S3 RawNet system performs best for the worst case A17 attack, it is of interest to evaluate the benefit of their fusion.

Experiments were conducted using the support vector machine (SVM) based fusion approach described in~\cite{IS2020LFCC}. It was used to fuse the high-spectral-resolution countermeasure (L) with S1, S2 and S3 RawNet2 systems. Fusion results are illustrated in Table~\ref{Tab:comparsion results}, with the top-4 ASVspoof 2019 challenge results.
Also shown are results for the two official ASVspoof 2019 baseline systems: the CQCC B1 baseline and the low-spectral-resolution LFCC B2 baseline.

All systems outperform the official baselines, with the high-spectral-resolution baseline (L) giving a min t-DCF of 0.0904, marginally better than the min t-DCF result of 0.0953 for team T24. Team T60 produced a min t-DCF of 0.0755 and team T45 a result of 0.0510. All fusions of the high-spectral-resolution countermeasure (L) with either S1, S2 and S3 RawNet2 systems perform better, with the L+S1 combination even outperforming fusion with all three RawNet systems (L+S1+S2+S3). \\

The last column of Table~\ref{Tab:comparsion results} shows performance in terms of min t-DCF for each countermeasure system for attack A17. Just like individual system results shown in Table~\ref{Tab: results on dev and eval}, all RawNet2 fused systems produce a substantially lower min t-DCF than the baseline system (L), confirming that the end-to-end RawNet2 system is complementary; it is learning artefacts that the baseline system is not. 
Last, while the results for team T05 show a lower pooled min t-DCF and lower min t-DCF for attack A17, our results are reproducible with open source software\footnote{https://github.com/eurecom-asp/rawnet2-antispoofing.}.

\section{Discussion and conclusions}
\label{sec:Conclusion}

This paper reports the first successful application of RawNet2 to automatic speaker verification anti-spoofing. While pooled results are inferior to those of the baseline countermeasure, those for the infamous, worst case A17 voice conversion spoofing attack are the second best reported thus far. Whether the comparison of countermeasure solutions should be based upon pooled or worst case results is an interesting question since, once a vulnerability is exposed, then attacks that exploit it will likely occur more frequently than less effective attacks. Despite the competitiveness of our proposed solution, the min t-DCF for attack A17 is still over 250\% higher than for the pooled min t-DCF.

The fusion of baseline and RawNet2 classifier scores give the second best results reported in the literature.
Fusion results suggest that the RawNet2 classifier is learning cues that are different to those learned by the baseline classifier. This observation raises the questions of what these cues are and what about the RawNet2 classifier enables it to learn them. Our current hypothesis is that, aside from operating upon the raw signal, the ability comes from temporal attention, an ability which the baseline classifier clearly lacks. 
This hypothesis would seem to be supported by the presence of occasional, punctual clicking noises that characterise A17 attacks. 
Our suspicion is that A17 artefacts are phase-related and that they are captured by the linear-phase filters in the first layer of the RawNet2 architecture which hence produce phase-aligned waveforms to the first residual block.
The validation of this hypothesis is the subject of ongoing work. Other directions for future work concern the exploration of embeddings and alternative back-end approaches to classification.

\vfill\pagebreak

\small

\bibliographystyle{IEEEbib}
\bibliography{strings,refs}

\end{document}

%% file: introNew.tex
Our recent efforts to improve the reliability of spoofing countermeasures for automatic speaker verification (ASV) have confirmed that: (i)~artefacts from converted voice or synthetic speech spoofing attacks reside at the sub-band level~\cite{odyssey2020CQCC}; (ii)~detection performance can be improved through the use of high-spectral-resolution front-ends, particularly those which emphasize information within the same sub-bands~\cite{IS2020LFCC}. When used even with a trivial Gaussian mixture model (GMM) back-end, high-spectral-resolution front-ends lead to competitive performance that would have obtained fourth position (out of 48 submissions) in the ASVspoof 2019 challenge. 

Unfortunately, though, some spoofing attacks continue to evade detection, particularly the infamous A17 attack, a neural network voice conversion (VC) based attack that uses a generalised direct waveform modification method for waveform generation. 
While we have seen that even A17 attacks can be detected when classifiers are learned with representative data (i.e.\ through learning with evaluation data), performance is poor when the same classifier is learned using only standard training data. Hence, while high-spectral-resolution front-ends \emph{do} capture A17 artefacts, they \emph{cannot} be detected.

The same observation shows that, once training data for a newly discovered spoofing attack is available, then vulnerabilities can be fixed easily simply by re-learning with representative training data. 
More proactive or preemptive strategies that offer some protection against the unexpected are preferable. This idea embodies the objectives of the ASVspoof initiative since its inception, namely the pursuit of generalisable spoofing countermeasures that are capable of detecting attacks not seen in training data. The latter are characteristic of in-the-wild settings in which the nature of spoofing attacks cannot be predicted and will evolve continuously. The question addressed in this paper is whether A17 attacks can be detected even in the \emph{absence} of representative training data.

This is a difficult question to answer reliably in a post-evaluation setting where the characteristics of attack A17 are now known and cannot be forgotten. One-class classifiers trained only on bona fide data without the use of any spoofed data are an obvious candidate solution and have shown some previous success in anomaly, novelty and spoofing detection tasks~\cite{EURECOM+4093,fatemifar2019spoofing,fatemifar2019combining,ohki2019efficient}. Anecdotal evidence, however, shows that many, including the authors of the current article, have failed to apply them successfully to ASVspoof 2019 data. Since A17 attacks cannot be detected using our current countermeasure (at least not without representative A17 training data) we must seek cues elsewhere. 

Our assumption is that the use of hand-crafted features does not offer the best potential to detect unforeseen attacks since they rely too much upon the characterisation of artefacts corresponding to known attacks. Attack-specific artefacts are likely to be insufficient for the detection of unseen attacks and a higher-level, more generalisable representation is needed to ensure robust performance. This paper reports what is, to the best of our knowledge, the first application of RawNet2~\cite{jung2020improved} deep neural network classifiers to anti-spoofing. RawNet2 operates directly upon the raw speech waveform, circumventing almost entirely any dependence upon hand-crafted features. The goal of this work is not just to design more reliable spoofing countermeasures, but also to determine whether such end-to-end approaches that utilise some degree of automatic feature learning can improve performance in a \emph{worst-case scenario} such as that of A17.

The paper is organised as follows. The original RawNet2 architecture and its application to ASV are described in Section~2. Modifications made to RawNet2 in order to apply it successfully to anti-spoofing are described in Section~3. Experimental work is described in Section~4. Our conclusions are presented in Section~5.